\newcommand\arcmin{\mbox{$^\prime$}}%
\documentclass[dvips]{article}
\usepackage{icrctc07,amsfonts,amssymb,amsmath}
\title{Observation of the Crab Nebula with the MAGIC telescope}
\shorttitle{Observation of the Crab Nebula with the MAGIC
telescope}
\authors{A. N. Otte$^{1,2}$, E. Aliu${^3}$, J. L. Contreras${^4}$,  M. Gaug$^{5}$,  M. Lopez$^{4}$, P. Majumdar$^{1}$\\ for the MAGIC collaboration}
\shortauthors{A. N. Otte and et al}
\afiliations{ $^1$Max-Planck-Institut f\"ur Physik, D-80805
M\"unchen, Germany\\ $^2$ Humboldt-Universit\"at zu Berlin,
D-12489 Berlin, Germany\\
$^3$ Institut de F\'\i sica d'Altes Energies, Edifici Cn., E-08193
Bellaterra (Barcelona), Spain\\
 $^4$
Universidad Complutense, E-28040 Madrid, Spain\\
$^5$ Universit\`a di Padova and INFN, I-35131 Padova, Italy}
\email{otte@mppmu.mpg.de}

\abstract{We report about very high energy (VHE) $\gamma$-ray
observations of
 the Crab Nebula with the MAGIC
 telescope. The $\gamma$-ray flux from the nebula was measured  between
 60\,GeV and 9\,TeV. The energy spectrum can be described with a
 curved power law $\frac{\mathrm{d}F}{\mathrm{d}E}=f_0\,
 \left(E/300\,\mathrm{GeV}\right)^{\left(a+b\log_{10}\left(E/300\,\mathrm{GeV}\right)\right)}
 $
 with a flux normalization $f_0$ of
 $(6.0\pm0.2_{\mathrm{stat}})\times10^{-10}\,$cm$^{-2}$s$^{-1}$TeV$^{-1}$,
 $a=-2.31\pm0.06_{\mathrm{stat}}$ and
 $b=-0.26\pm0.07_{\mathrm{stat}}$. The position of the IC-peak is determined at $(77\pm47)\,$GeV. Within the observation time and the experimental
 resolution of the telescope, the $\gamma$-ray emission is steady and pointlike.
 The emission's center of gravity
 coincides with the position of the pulsar. Pulsed $\gamma$-ray emission from the pulsar could not be detected.
 We constrain the cutoff energy of the spectrum to
 be less than $\sim30$\,GeV, assuming that the differential
 energy spectrum has an exponential cutoff. For a super-exponential shape, the cutoff
 energy can be as high as $\sim60$\,GeV.}

\begin{document}
\maketitle

\section{Introduction}

The Crab Nebula is the remnant of a supernova explosion that
occurred in 1054 A.D.~at a distance of $\sim2\,$kpc.
It is one of the best studied non-thermal celestial objects in
almost all wavelength bands of the electromagnetic spectrum from
$10^{-5}\,$eV (radio) to nearly $10^{14}$\,eV ($\gamma$-rays). The
radiation from radio to $\gamma$-rays (E $\le$ 1 GeV) is
interpreted as synchrotron emission of relativistic electrons and
positrons. At higher energies it is believed that inverse Compton
scattering is the dominant generation process of $\gamma$-rays
\cite{1965PhRvL..15..577G,1992ApJ...396..161D,1989ApJ...342..379W}.
There is little doubt that the engine of the nebula is the pulsar
PSR B0531+21 (hereafter Crab pulsar), which is also a strong
source of $\gamma$-rays detected up to 10\,GeV.

In very high energy (VHE) $\gamma$-ray astronomy the Crab nebula
was first detected with large significance at TeV energies by the
pioneering Whipple telescope \cite{1989ApJ...342..379W}. Since
then the Crab nebula was extensively studied by ground based
experiments at energies above a few hundred GeV. However, between
$10\,$GeV and $\sim200\,$GeV, observations are sparse.

Here we present highlights of an 16 hour observation of the Crab
nebula and pulsar that was performed with the MAGIC telescope
between October 2005 and December 2005. A more detailed discussion
of the analysis and results presented here can be found in
\cite{magiccrab}.

After a short description of the MAGIC telescope and the performed
observations we present results from the analysis of the
VHE-$\gamma$-ray emission from the nebula and the search for
pulsed emission from the pulsar. The paper is closed with some
concluding remarks.

\section{The MAGIC telescope}

The MAGIC (Major Atmospheric Gamma Imaging Cherenkov) telescope is
located on the Canary Island La Palma (2200 m asl, $28.45^\circ$N,
$17.54^\circ$W). MAGIC is currently the largest single dish (17\,m
diameter) imaging atmospheric Cherenkov telescope (IACT). The
faint Cherenkov light flashes produced in air showers are recorded
by a camera comprising 577 photomultiplier tubes (PMTs). The
central PMT is modified for optical pulsar studies
\cite{2005ICRC....5..367L}.

The current configuration of the MAGIC camera has a trigger region
of 2.0 degrees in diameter \cite{2005ICRC....5..359C}. 
Presently, the trigger energy range spans from 50-60 GeV (at small
zenith angles) up to tens of TeV.

\section{Analysis results of the Crab Nebula}

\subsection{Source Morphology}

\begin{figure}[htb]
  \begin{center}
\includegraphics*[width=0.47\textwidth]{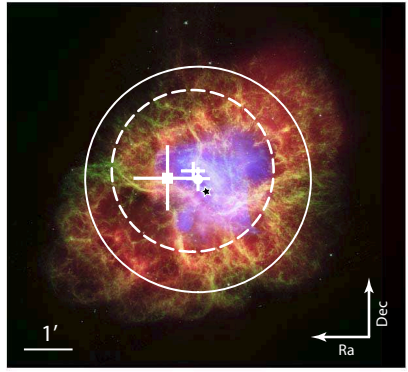}
\end{center}

\caption{The crosses indicate the CoG of the VHE $\gamma$-ray
emission at different energies (simple cross $>500\,$GeV; dot
$\sim250\,$GeV; square $\sim160\,$GeV), overlaid on an optical
image from HST and an X-Ray image from Chandra. The position of
the Crab pulsar is marked with a black star. The error bars
indicate the statistical uncertainty in the position of the CoG.
The upper limits  on the 39\% containment radius of the
$\gamma$-ray emission region are indicated with circles (dashed
$>500\,$GeV; solid $\sim250\,$GeV).} \label{crabcomposite}
  \end{figure}

The morphology of the $\gamma$-ray emission was studied by
generating sky-maps in three different bins of energy. The center
of gravity (CoG) of the $\gamma$-ray emission was derived from the
sky-maps  by fitting them with a 2D-Gaussian of the form
\begin{equation}\label{2dgaus}
    F_{\mathrm{res}}+a\cdot\exp\left[-\frac{(x-\bar{x})^2+(y-\bar{y})^2}{2\sigma^2}\right],
\end{equation}
where $F_{\mathrm{res}}$ is introduced to account for a possible
constant offset of the background subtracted sky-map. The CoGs
obtained from the fitted $\bar{x}$ and $\bar{y}$ are  shown in
Figure \ref{crabcomposite} superimposed on the composite image of
optical, IR and X-ray observations of the Crab nebula. The three
measured CoGs are compatible among each other and coincide with
the position of the pulsar. Note that the systematic uncertainty
of the position is about $1'$.

The extension of the $\gamma$-ray emission region is compatible
with a point-like source. Upper limits on the source extension
were calculated with a confidence level of 95\%. The results  for
energies $\sim250$\,GeV ($<2.4\arcmin$) and $>500\,$GeV
($<1.6\arcmin$) are presented in Figure \ref{crabcomposite}.  In
both cases the emission is constrained to originate from within
the optical nebula.

\subsection{Energy spectrum}

\begin{figure}
\begin{center}
\includegraphics*[width=\columnwidth]{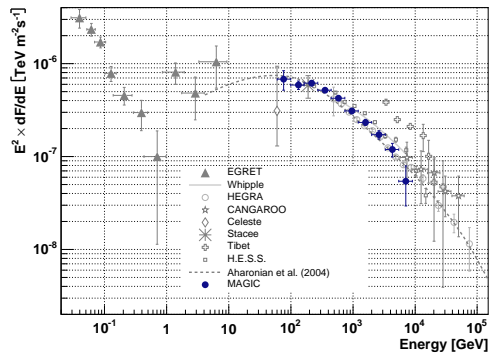}
\end{center}
\caption{Spectral energy distribution of the $\gamma$-ray emission
of Crab nebula. Below 10\,GeV, measurements are by EGRET.
                        In VHE $\gamma$-rays measurements
                        are from ground-based experiments. The dashed
                        line is a model
                        prediction by \cite{2004ApJ...614..897A}.
                        }
                        \label{Crab_sed_time}
\end{figure}

Figure \ref{Crab_sed_time} shows the differential flux
measurements by MAGIC multiplied by the energy squared, i.e.~the
spectral energy density distribution. A parameterization of the
spectrum with a power-law ansatz results in a $\chi^2$ of 24 for 8
degrees of freedom. A better parameterization is obtained with a
curved power-law ansatz.
\begin{equation}
\frac{\mathrm{d}F}{\mathrm{d}E}=f_0\,
\left(E/300\,\mathrm{GeV}\right)^{\left(a+b\log_{10}\left(E/300\,\mathrm{GeV}\right)\right)}
\end{equation}
yielding a flux normalization $f_0$ of
$(6.0\pm0.2_{\mathrm{stat}})\times10^{-10}\,$cm$^{-2}$s$^{-1}$TeV$^{-1}$,
$a=-2.31\pm0.06_{\mathrm{stat}}$ and
$b=-0.26\pm0.07_{\mathrm{stat}}\pm0.2_{\mathrm{syst}}$. The
$\chi^2$ of the fit is 8 for 7 degrees of freedom.

For energies above 400\,GeV the derived spectrum is in good
agreement with measurements of other air Cherenkov telescopes. At
energies $<400\,$GeV, below the threshold of previous measurements
by IACTs, we compare our results with integral flux measurements
obtained by \cite{2002ApJ...566..343D} and
\cite{2001ApJ...547..949O}.

\begin{figure}[htb]
\begin{center}

                    \includegraphics*[width=0.47\textwidth]{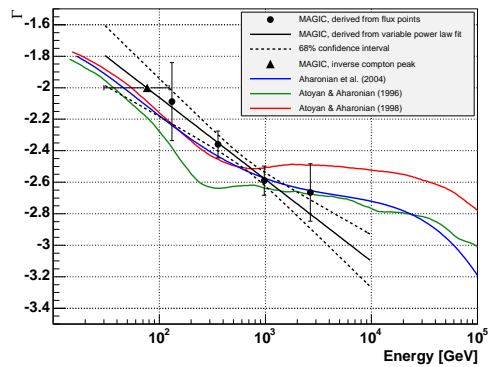}
\end{center}
                        \caption{Spectral index derived from differential flux measurements (dots) and from the curved power-law
                        fit (black solid line, the dashed line gives the $1\sigma$ confidence interval); Predictions
                        by \cite{2004ApJ...614..897A} (blue curve),
                        \cite{1996MNRAS.278..525A} (green curve) and
                        \cite{1998nspt.conf..439A} (red curve) are also shown.
                        }
                        \label{photindex}
  \end{figure}

At lower energies one expects a continuous softening of the
spectrum with increasing energy. However, this could not be
demonstrated by earlier measurements. We derived spectral indices
at $\sim150\,$GeV, $\sim300\,$GeV, $\sim1\,$TeV and $\sim2.5\,$TeV
from the flux measurements as well as from the aforementioned
results of the curved power-law fit. The results, shown in Figure
\ref{photindex} together with several predictions, indicate a
clear softening of the spectrum with
increasing energy. 

The predicted GeV $\gamma$-ray emission has a peak in the
SED-representation (see Figure \ref{Crab_sed_time}). If one
assumes that the energy spectrum around the peak can be described
with a curved power-law, we can determine the position of the peak
from the curved power-law fit to be at
$77\pm47_\mathrm{stat}{\genfrac{}{}{0pt}{2}{+107}{-46}}_\mathrm{syst}$\,GeV.

Within statistical uncertainties, the flux of $\gamma$-rays was
constant over the entire observation period. Tested timescales
were between a few minutes up to months.
The average integral flux above 200\,GeV is
$(1.96\pm0.05_{\mathrm{stat}})\times
10^{-10}\,\mathrm{cm}^{-2}\,\mbox{s}^{-1}$.

\section{VHE-$\gamma$-ray emission from the pulsar}

 \begin{figure}[htb]
        \begin{center}
\includegraphics*[width=0.47\textwidth]{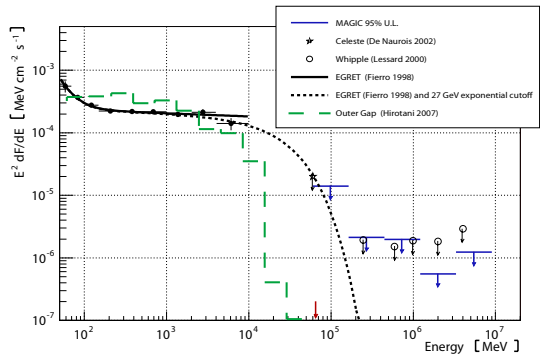}
\end{center}
                        \caption{Upper limits on the pulsed gamma ray flux from the Crab
                        pulsar; upper limits in differential bins of energy are given
                        by the blue points. The upper limit on the
                        cutoff energy of the pulsed emission is
                        indicated by the dashed line.}
                        \label{Crab_pulsed_differential}
    \end{figure}

Motivated by a possible pulsed $\gamma$-ray component at TeV
energies \cite{2001ApJ...549..495H}, we searched for pulsed
emission in five bins of reconstructed energy between $60\,$GeV
and $9\,$TeV. However, a signature of periodicity was not found in
any of the tested energy intervals. Derived 95\% confidence level
flux limits are shown in Figure \ref{Crab_pulsed_differential}.

\begin{figure}[htb]
   \begin{center}
\includegraphics*[width=0.47\textwidth]{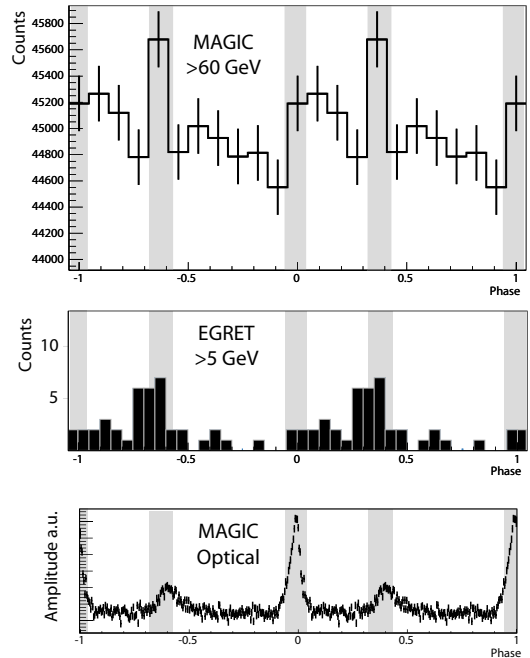}
\end{center}
\caption{\label{lcoptcuts}Pulse phase profiles of the Crab pulsar.
Lower figure: optical observations by MAGIC; middle figure:
observations by EGRET above 5\,GeV; upper figure: pulse phase
profile obtained by MAGIC. The shaded regions indicate the EGRET
measured positions of the pulsed emission for $\gamma$-ray
energies above 100\,MeV.}
   \end{figure}

We also performed a periodicity analysis optimized for a search of
pulsed emission close to the threshold of the experiment (analysis
threshold 60\,GeV). Figure \ref{lcoptcuts} shows the resulting
pulse phase profile together with EGRET data above 5\,GeV and
measurements in optical by MAGIC. An excess in VHE-$\gamma$-rays
is evident at the position of the inter-pulse in the same phase
range where EGRET detected pulsed emission above 100\,MeV (shaded
region) and above 5\,GeV. The MAGIC and EGRET $>5\,$GeV pulse
phase profile match with a probability of 87\%. One calculates a
significance of $2.9\,\sigma$ of the excess if the phase regions
where EGRET detected pulsed emission above 100\,MeV (shaded
regions) are defined as signal region and the remaining phase
intervals as background region.

The observed excess is not sufficient to claim the detection of a
pulsed signal, therefore, upper limits on the number of excess
events where calculated with a confidence level of 95\%. Using the
limit on the number of pulsed excess events we constrained  the
cutoff energy of the pulsar spectrum to be less than $27$\,GeV,
under the assumption that the break in the energy spectrum can be
described with an exponential cutoff. In case the energy spectrum
is attenuated super-exponentially, cutoff energies up to
$\sim60$\,GeV are allowed from our observations.

\section{Concluding remarks}

Here we reported on the currently most detailed study of the VHE
$\gamma$-ray emission of the Crab nebula below 500\,GeV. Most of
the aforementioned studies were done in this energy region for the
first time. Results from this study among others are a:
   \begin{itemize}
      \item{measurement of the differential energy spectrum down to 60\,GeV, which clearly deviates from a pure power-law behavior}
      \item{determination of the inverse Compton peak at $77\pm47_\mathrm{stat}{\genfrac{}{}{0pt}{2}{+107}{-46}}_\mathrm{syst}$\,GeV}
      \item{point-like emission region with a CoG coinciding with the position of the pulsar }
      \item{constraint of the cutoff energy of
      the pulsar spectrum of $\lesssim27\,$GeV, assuming an exponential cutoff}
   \end{itemize}

\section{Acknowledgements}
We are grateful for discussions with Kouichi Hirotani. We also
would like to thank the IAC for the excellent working conditions
at the ORM in La Palma. The support of the German BMBF and MPG,
the Italian INFN, the Spanish CICYT, ETH research grant TH 34/04
3, and the Polish MNiI grant 1P03D01028 are gratefully
acknowledged.
\bibliography{icrc1078}
\bibliographystyle{plain}

\end{document}